\documentclass[11pt]{article}
\usepackage{epsfig}
\usepackage{graphicx}
\usepackage{amssymb}
\usepackage{here}
\usepackage{cite}
\newcommand{\AmS}{{\protect\the\textfont2
  A\kern-.1667em\lower.5ex\hbox{M}\kern-.125emS}}

\newcommand{\lsim}{\raisebox{-0.13cm}{~\shortstack{$<$ \\[-0.07cm] $\sim$}}~} 
\newcommand{\gsim}{\raisebox{-0.13cm}{~\shortstack{$>$ \\[-0.07cm] $\sim$}}~} 
\def\beq{\begin{equation}}   
\def\eeq{\end{equation}}

\def\nn{\nonumber}
\def\bea{\begin{eqnarray}}
\def\eea{\end{eqnarray}}
\def\Beff{B_{\rm eff}}
\def\mueff{\mu_{\rm eff}}

\begin{document}
\begin{titlepage}

\begin{flushright}
Report No: IFIC/08-53, FTUV-08-1024, LPT-08-81, \\
ANL-HEP-PR-08-64, LPTA/08-059\\
\end{flushright}

\begin{center}
\vspace{3cm}
{\Large{\bf Radiative $\Upsilon$ decays and a light
pseudoscalar Higgs in the NMSSM}}
\\
\vspace{1cm}

{\bf{Florian Domingo$^a$, Ulrich Ellwanger$^a$, Esteban
Fullana$^b$, Cyril Hugonie$^c$ and  Miguel-Angel Sanchis-Lozano$^d$
\vspace{2cm}\\
\it 
$^a$ Laboratoire de Physique Th\'eorique\footnote{Unit\'e mixte de
Recherche -- CNRS -- UMR 8627}, Universit\'e de Paris--Sud, F--91405
Orsay, France \\
$^b$ High Energy Physics Division, Argonne National Laboratory, 
Argonne, IL~60439, USA \\
\it $^c$ LPTA\footnote{Unit\'e mixte de Recherche -- CNRS -- UMR 5207},
Universit\'e de Montpellier II, 34095 Montpellier, France\\
\it $^d$ Instituto de F\'{\i}sica Corpuscular (IFIC)
and Departamento de F\'{\i}sica Te\'orica \\
\it Centro Mixto Universitat de Val\`encia-CSIC, 
Dr. Moliner 50, E-46100 Burjassot, Valencia, Spain}}

\begin{abstract}
We study possible effects of a light CP-odd Higgs boson on radiative
$\Upsilon$ decays in the Next-to-Minimal Supersymmetric Standard Model.
Recent constraints from CLEO on radiative $\Upsilon(1S)$ decays are
translated into constraints on the parameter space of CP-odd Higgs
boson masses and couplings, and compared to constraints from $B$ physics
and the muon
anomalous magnetic moment. Possible Higgs-$\eta_b(nS)$ mixing effects
are discussed, notably in the light of the recent measurement of the
$\eta_b(1S)$ mass by Babar: The somewhat large $\Upsilon(1S)$ -
$\eta_b(1S)$ hyperfine splitting could easily be explained by the
presence of a CP-odd Higgs boson with a mass in the range 9.4 -
10.5~GeV. Then, tests of lepton universality in inclusive radiative
$\Upsilon$ decays can provide a visible signal in forthcoming
experimental data. 
\end{abstract}


\end{center}
\end{titlepage}


\section{Introduction}

The Next-to-Minimal Supersymmetric Standard Model (NMSSM)~\cite{nmssm}
provides the simplest solution to the $\mu$~problem of the
MSSM~\cite{muprob}. Its phenomenology can differ in various respects
from the MSSM. Notably, as emphasized in~\cite{dobr}, a light CP-odd
scalar $A_1$ can appear in the Higgs spectrum. 

In the case where the CP-even Higgs boson $H$
decays dominantly into a pair of CP-odd
scalars~\cite{dobr,hsearch1,dg1,Dermisek:2005gg,dg2,hsearch2,Dermisek:2007yt,hsearch3},
LEP constraints on CP-even Higgs masses~\cite{Schael:2006cr} are
alleviated
considerably~\cite{dg1,Dermisek:2005gg,dg2,hsearch2,Dermisek:2007yt,hsearch3}.
For $m_{A_1} \lsim$ 10.5~GeV, where the $A_1$ decay into $B\bar{B}$ is
forbidden, this scenario could even explain the $2.3\, \sigma$ excess in
the $e^+ e^- \to Z + 2b$ channel for $M_{2b} \sim
100$~GeV~\cite{Dermisek:2005gg} (where the two $b$ quarks would result
from a CP-even Higgs $H$ with $M_H \sim 100$~GeV and a branching ratio
${\cal B} \left(H
\to b\bar{b}\right) \sim 0.08$, but ${\cal B} \left(H
\to A_1 A_1\right) \sim 0.9$). Also at hadron colliders
the search for CP-even scalars would be particularly
difficult~\cite{hsearch1, Dermisek:2005gg,hsearch2,
Dermisek:2007yt,hsearch3} if they decay dominantly into $A_1 A_1$ with
$m_{A_1}$ below the $B\bar{B}$ threshold. In this case,
however, the $A_1$ can have important effects on
$\Upsilon$~decays~\cite{Drees:1989du, SanchisLozano:2002pm,McElrath:2005bp,
Sanchis-Lozano:2006gx, Dermisek:2006py,Fullana:2007uq,
Hodgkinson:2008ei}. Notably a Super B factory can then play an important
and complementary role~\cite{Bona:2007qt} via its potential sensitivity
to $\Upsilon \to \gamma A_1$ decays.

Whereas a light $A_1$ is also possible in the MSSM with a CP-violating
Higgs sector~\cite{Carena:2002bb,Lee:2007ai}, scenarios with more than
one gauge singlet~\cite{Han:2004yd}, little Higgs models and
non-supersymmetric two Higgs doublet models (see~\cite{Kraml:2006ga} for
an overview), we concentrate subsequently on the simplest version of the
NMSSM with a scale invariant superpotential. 

$\Upsilon \to \gamma A_1$ decays have been investigated in the
NMSSM before in~\cite{Dermisek:2006py} for $m_{A_1} \lsim 9.2$~GeV, where the
signal relies on a narrow peak in the photon spectrum. Recent results
from CLEO on radiative $\Upsilon(1S)$ decays~\cite{cleo} (assuming a
$A_1$ width $\lsim 10$~MeV) constrain this domain of $m_{A_1}$ strongly.
One of our aims is to translate the CLEO results into constraints on the
NMSSM parameter space (see also~\cite{gunion}) as $X_d$, the
(reduced) coupling of $A_1$ to $b$ quarks, and to compare them with
constraints from B~physics~\cite{Hiller:2004ii, Domingo:2007dx,
Heng:2008rc} and the anomalous magnetic moment of the
muon~\cite{mumagmo,gunion}.

For $m_{A_1} \gsim 9$~GeV, various corrections to the $\Upsilon \to
\gamma A_1$ decay rate become relatively large and
uncertain~\cite{guide}, which makes it difficult to translate
experimental constraints on the decay rate into constraints on $X_d$.
Consequently, this region of $m_{A_1}$ is hardly constrained by CLEO
results.

Very recently, a CP-odd state with a mass of about 9389~MeV has been
observed in $\Upsilon (3S)$ decays by BaBar~\cite{Aubert:2008vj},
showing up as a peak with a significance of 10 standard deviations in
the photon energy spectrum. At first sight, this state can be
interpreted as the long-awaited $\eta_b(1S)$. However, in the presence
of a CP-odd Higgs with a mass in the same region, the observed mass has
to be interpreted as an eigenvalue of a $2 \times 2$ mixing matrix, and
would differ correspondingly from $m_{\eta_{b0}}$, the mass of the
$\eta_b(1S)$ in the absence of a CP-odd Higgs ($m_{\eta_{b0}}^2$ is now one
of the diagonal entries of the mass matrix)~\cite{Drees:1989du}. In this
case, a second peak in the photon spectrum should possibly be visible;
however, the search for such a second peak would require a dedicated
consideration of the various background contributions (notably from the
ISR and $\chi_{bJ}(2P)$), which should be performed in the future.

First, this mixing effect could explain the fact that the observed mass
is somewhat lower than expected, if $m_{A_1}$ is somewhat above
$m_{\eta_{b0}}$. Second, the off-diagonal element of the mass matrix can be
estimated and turns out to be proportional to $X_d$~\cite{Drees:1989du,
Fullana:2007uq}. Assuming a reasonable range for $m_{\eta_{b0}}$, the
observed value of $\sim$~9389~MeV for one of the eigenvalues implies an
upper bound on $X_d$ as a function of $m_{A_1}$, which is, however,
particularly strong only for $m_{A_1}\sim$~9389~MeV and will be derived
below.

At present, a direct detection of a CP-odd Higgs with a mass in the
particularly interesting region 9.2~GeV~$\lsim m_{A_1} \lsim 10.5$~GeV
via a peak in the photon spectrum seems to be quite difficult.
Fortunately, an alternative signal for an $A_1$ state below the $B
\bar{B}$ threshold can be a breakdown of lepton universality (LU) in
$\Upsilon \to (\gamma)\, l^+ l^-$ decays (via an intermediate $A_1$
state), since $A_1$ would decay practically exclusively into $\tau^+
\tau^-$~\cite{SanchisLozano:2002pm,Sanchis-Lozano:2006gx, Fullana:2007uq}. Note
that, to this end, the photon does not have to be detected. Present
tests of lepton universality in $\Upsilon(1S)$, $\Upsilon(2S)$ and 
$\Upsilon(3S)$ decays~\cite{Amsler:2008zz} have error bars in the
5--10\% range. Remarkably, however, a general trend (at the
$1\,\sigma$~level) seems to point towards a slight excess of the $\tau^+
\tau^-$ branching ratios, as expected in the presence of a $A_1$ state.
Another aim of the present paper is to  investigate corresponding
sensitivities of forthcoming experimental data, assuming a possible
reduction of the errors to the 2\% range.

The layout of this paper is as follows: In section~\ref{sec:nmssm} we
review the domains of the NMSSM parameter space which lead to a light
CP-odd Higgs with strong couplings to down type quarks (and leptons). In
section~\ref{sec:cleo} we derive constraints on the NMSSM parameter
space from recent CLEO results, using quite conservative estimates for
the corrections to the $\Upsilon \to \gamma A_1$ decay rate which lead to
quite conservative upper bounds on $X_d$ as a function of $m_{A_1}$.
(These upper bounds on $X_d$ will be included in future versions of the
code NMSSMTools~\cite{nmssmtools}.) In section~\ref{sec:mix}, we discuss
the mixing effects of $A_1$ with $\eta_b(nS)$
following~\cite{Drees:1989du,SanchisLozano:2002pm,Fullana:2007uq}. In
section~\ref{sec:babar}, we derive constraints on $X_d$ from the
measured $\eta_{b_{obs}}$ mass by BaBar and from (conservative)
assumptions on $m_{\eta_{b0}}$, and discuss quantitatively the possible
mixing-induced shift of the measured $\eta_{b_{obs}}$ mass. In
section~\ref{sec:comp} we compare these CLEO and BaBar constraints with
constraints from LEP, B~physics and the muon anomalous magnetic moment.
This analysis is performed with the help of the updated NMSSMTools
package. In section~\ref{sec:luv} we reconsider $A_1$ masses between 9.2
and 10.5~GeV and show that (for less conservative estimates of the
corrections to the $\Upsilon \to \gamma A_1$ decay rate) a breakdown of
lepton universality in $\Upsilon \to (\gamma)\, l^+ l^-$ decays can
become an important observable for the detection of a CP-odd Higgs in
this mass range. We present formulas for the relevant branching ratios
including possible $A_1 - \eta_b(nS)$ mixings, and study future
sensitivities on $X_d$ from lepton universality breaking.
Section~\ref{sec:concl} contains conclusions and an outlook.

\section{A light CP-odd Higgs in the NMSSM}\label{sec:nmssm}

In this section we show that the parameter space of the NMSSM can
accomodate a light CP-odd Higgs, which is strongly coupled to
down-quarks and leptons (see also~\cite{dobr,Dermisek:2005gg,dg2}). We
consider the simplest version of the NMSSM with a scale invariant
superpotential
\beq
W= \lambda S H_u H_d +\frac{1}{3}\kappa S^3 +\dots
\eeq
and associated soft trilinear couplings
\beq
V_{soft}=(\lambda A_\lambda S H_u H_d +\frac{1}{3}\kappa A_\kappa S^3)
+h.c. +\dots 
\eeq
in the conventions of~\cite{nmssmtools}. A vev of the singlet
field $s \equiv \left<S\right>$ generates an effective $\mu$-term, and
it is convenient to define also an effective $B$-term:
\beq
\mueff=\lambda s,\qquad \Beff=A_\lambda + \kappa s\; .
\eeq

The Higgs sector of the NMSSM contains six independent parameters, which
can be chosen as
\beq\label{eq:inp}
\lambda,\ \kappa,\ A_\lambda,\ A_\kappa,\ \tan\beta,\ \mueff \; .
\eeq

In the NMSSM, two physical pseudoscalar states appear in the spectrum,
which are superpositions of the MSSM-like state $A_{MSSM}$ (the
remaining $SU(2)$ doublet after omitting the Goldstone boson) and the
singlet-like state $A_S$. In the basis $(A_{MSSM},\, A_S)$, the $2
\times 2$ mass square matrix for the CP-odd Higgs bosons has the
following matrix elements~\cite{nmssmtools}
\begin{eqnarray}
M_{11}^2&=&\frac{2 \mueff \Beff}{\sin 2\beta},\qquad
M_{12}^2\ =\  \lambda v (A_{\lambda}-2 \kappa s) \nn \\ 
M_{22}^2&=& \frac{\lambda^2 v^2 \sin 2\beta}{2 \mueff} \left(A_\lambda
+4\kappa s\right)-3 \kappa s A_{\kappa}
\end{eqnarray}
where $v^2 = 1/(2\sqrt{2} G_F)$.  The masses of the CP-odd eigenstates
$A_{1,2}$ are
\begin{equation} 
m_{A_{1,2}}^2=\frac{1}{2}[M_{11}^2+M_{22}^2 \mp \Delta M^2]
\label{eq:mA12} 
\end{equation}
with $\Delta M^2=\sqrt{(M_{11}^2-M_{22}^2)^2+4(M_{12}^2)^2}$.

The lighter CP-odd state $A_1$ can be decomposed into $(A_{MSSM},\,
A_S)$ according to 
\begin{equation} 
A_1=\cos\theta_A A_{MSSM}+\sin\theta_A A_S\; ,
\end{equation} 
where the mixing angle $\theta_A$ is
\begin{equation}
\cos 2\theta_A=\frac{M_{22}^2-M_{11}^2}{\Delta M^2}\; .
\end{equation}

To a good approximation (for moderate $A_\lambda$, small $A_\kappa$ and
large $\tan\beta$), the mass of the lightest CP-odd Higgs boson and
$\cos\theta_A$ can be written as~\cite{dobr,dg2}
\begin{equation}
m_{A_1}^2 \simeq 3 \kappa s \left(\frac{3\lambda^2 v^2 A_\lambda\sin 2\beta}
{2\mueff\Beff -3\kappa s A_\kappa\sin 2\beta} -A_\kappa
\right)\; ,
\label{eq:mass}
\end{equation}
\begin{equation}
\cos\theta_A \simeq -\frac{\lambda v(A_{\lambda}-2 \kappa
s)\sin 2\beta} {2\mueff\Beff+3 \kappa s A_{\kappa} \sin 2\beta}\; .
\label{eq:cos} 
\end{equation}
(The approximate equation for $\cos\theta_A$ ceases to be valid if the
second term in the denominator is large compared to the first one.)

The reduced coupling $X_d$ of the light physical $A_1$ Higgs boson to
down-type quarks and leptons (normalized with respect to the coupling
of the CP-even Higgs boson of the Standard Model) is given~by
\begin{equation}\label{eq:xd}
X_d=\cos\theta_A\tan\beta\; .
\end{equation}

Interesting phenomena in the $\Upsilon$-system take place for large
values of $X_d$, i.e. large values of $\tan\beta$ without
$\cos\theta_A$ being too small. (A possible enhancement of
$X_d$~\cite{Hodgkinson:2008ei} can occur due to the radiatively
generated $\tan\beta$-enhanced Higgs-singlet Yukawa
couplings~\cite{Hodgkinson:2006yh}.  How\-ever, in the case of a sizable
value of $\cos\theta_A$ already at tree level as considered below, this
effect is small.) 

At first sight eq. (\ref{eq:cos}) seems to imply (from $\sin 2\beta
\sim 2/\tan\beta$ for large $\tan\beta$) that $\cos\theta_A$ decreases
indeed with $\tan\beta$ -- this would be the case in the
PQ-symmetry-limit ($\kappa \to 0$) or R-symmetry-limit ($A_\kappa,\
A_\lambda \to 0$), where the second term in the denominator of
(\ref{eq:cos}) tends to zero. On the other hand, it follows from the
minimization equations of the scalar potential of the NMSSM (as in the
MSSM), for fixed soft Higgs mass terms and $\mueff$, that $\tan\beta$ is
proportional to $1/|\mueff \Beff|$ for large
$\tan\beta$~\cite{Ananthanarayan:1996zv}, hence large values of
$\tan\beta$ are associated to small values of $|\Beff|$ (since $|\mueff|
\gsim 100$~GeV from the lower bound on chargino masses). It is useful to
replace $|\Beff|$ by the parameter 
\beq\label{eq:ma}
M_A^2 \equiv M_{11}^2 = \frac{2 \mueff \Beff}{\sin 2\beta}\; ,
\eeq
which sets the scale for the masses of the complete $SU(2)$ multiplet of
Higgs states including a scalar, a pseudoscalar and a charged Higgs as
in the MSSM (in our case, the corresponding pseudoscalar is the heavier
one $A_2$). In terms of $M_A^2$, $X_d$ can be written approximately as
\begin{equation}
X_d \simeq -\frac{\lambda v(A_{\lambda}-2 \kappa s)}
{M_A^2+3 \kappa s A_{\kappa}} \times \tan\beta\; ,
\label{eq:XdMA}
\end{equation}
and it is reasonable to examine the large $\tan\beta$ region keeping
$M_A$ fixed.

It follows from eq. (\ref{eq:mass}) that there exist always values of
$A_\kappa$ of the same sign as $A_\lambda$ (typically
both negative) where $m_{A_1}$ is small~\cite{Dermisek:2007yt}, while
$\cos\theta_A\simeq 0.1-0.6$ and hence $X_d$ is not suppressed. This
requires a moderate fine-tuning of $A_\kappa$ (or $M_A$); on the other
hand the authors of Ref.\cite{Dermisek:2006py} stress that such values
for $A_{\kappa}$, which allow a light SM-like Higgs to decay into two
$A_1$ with $m_{A_1} < 2m_b$, correspond to the smallest degree of
fine-tuning in the entire parameter space of the NMSSM. For
corresponding values of $A_\kappa$, the denominator of $X_d$ in
(\ref{eq:XdMA}) is dominated by $M_A^2$. For $M_A^2 \lsim |\kappa
A_\kappa s|$, even larger values of $\cos\theta_A\simeq 0.6-1.0$ are
possible while $m_{A_1}$ remains small. In this regime, the
approximations leading to eqs. (\ref{eq:mass}), (\ref{eq:cos}) and
(\ref{eq:XdMA}) are no longer valid, however.

To summarize, the following conditions can be fulfilled simultaneously
in the NMSSM, which yield possibly observable effects in $\Upsilon$
decays:
\begin{itemize}
\item $m_{A_1} \lsim 10.5$~GeV from, e.g., appropriate values of
$A_\kappa$;
\item a large value of $X_d$, if $\tan\beta$ is large while $M_A$
in the denominator of (\ref{eq:XdMA}) remains moderate.
\end{itemize}
The numerical results in section~\ref{sec:comp} confirm the analytical
estimates above.

\section{Constraints from CLEO}\label{sec:cleo}

Recently, the CLEO collaboration presented results on Higgs searches
from $\Upsilon(1S)$ decays~\cite{cleo}. $21.5\cdot 10^6\ \Upsilon(1S)$
decays had been collected and, for the $\Upsilon(1S) \to \gamma + (A_1 
\to \tau^+\tau^-)$ search, the photon energy spectrum in events with
missing energy and one identified $\mu^\pm$ or $e^\pm$ (allegedly from
$\tau \to e\nu\nu$ or $\tau \to \mu\nu\nu$) had been examined. For the
$A_1 \to \mu^+\mu^-$ search, both muons were identified.

No narrow peaks (of a width below $\sim 10$~MeV) in the photon energy
spectrum are observed (except for $\Upsilon(1S) \to \gamma J/\Psi \to
\gamma\, \mu^+\mu^-$), which allows to place stringent upper limits
between $10^{-4}$ and $10^{-5}$ on the branching ratio ${\cal B}
\left(\Upsilon(1S) \to \gamma (A_1 \to \tau^+\tau^- /\mu^+\mu^-)\right)$
for $m_{A_1} \lsim 9.2$~GeV~\cite{cleo}.

The ${\cal B} \left(\Upsilon(1S) \to \gamma A_1\right)$ is given by the
Wilczek formula~\cite{Wilczek:1977pj,haber}
\beq\label{eq:wilczek}
\frac{{\cal B} \left(\Upsilon(1S) \to \gamma A_1\right)}
{{\cal B} \left(\Upsilon(1S) \to \mu^+\mu^-\right)} =
\frac{G_F m_b^2 X_d^2}
{\sqrt{2}\pi\alpha}\biggl(1-\frac{m_{A_1}^2}{m_{\Upsilon(1S)}^2}\biggr)
\times  F 
\eeq
where $\alpha$ denotes the fine structure constant and
$X_d$ is given in (\ref{eq:xd}). $F$ is a correction factor, which
includes three kinds of corrections to the leading-order Wilczek
formula (the relevant formulas are summarized in \cite{guide}): bound
state, QCD and relativistic corrections. Bound state effects have a
quite different behaviour for a scalar or a pseudoscalar Higgs,
increasing the ratio (\ref{eq:wilczek}) by $\sim 20\%$ in the latter
case~\cite{Polchinski:1984ag,Pantaleone:1984ug,Bernreuther:1985ja}. QCD
corrections reduce the ratio (\ref{eq:wilczek}) by a similar
amount~\cite{Vysotsky:1980cz,Nason:1986tr}. Relativistic corrections can
generate an important reduction, and were calculated
in~\cite{Aznaurian:1986hi}.

\begin{figure}[ht!]
\begin{center}
\includegraphics[scale=0.5,clip=,]{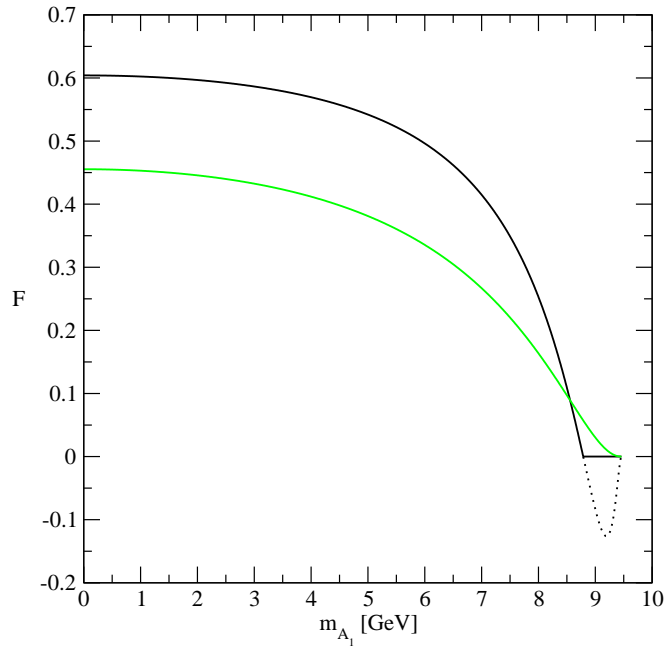}
\caption{$F(m_{A_1})$ including the bound state and QCD corrections,
and a naive extrapolation of the relativistic corrections computed for
$m_{A_1} \ll m_\Upsilon$ at larger values of $m_{A_1}$. Black curve:
assuming $m_b = 4.9$~GeV (as used later), green curve: assuming
$m_b = 5.3$~GeV.}
\label{fig:F}
\end{center}
\end{figure}

These relativistic corrections depend quite strongly on the $b$ quark
mass $m_b$, and become unreliable at least for Higgs masses $m_{A_1}$
above 8~GeV~\cite{Aznaurian:1986hi} where they can generate a vanishing
(or even negative) correction factor~$F$. Frequently, the approximation $F \sim
0.5$ for all $m_{A_1}$ is employed in the literature~\cite{haber,Hiller:2004ii}.
However, in order to derive conservative bounds on the NMSSM parameters
from CLEO results, we use in this section the smaller values of 
$F(m_{A_1})$ for larger $m_{A_1}$, which are obtained by a naive
extrapolation of the relativistic corrections~\cite{Aznaurian:1986hi}.
Using the quark model value $m_b = 4.9$~GeV, the resulting behaviour of
$F(m_{A_1})$ (including also the bound state and QCD corrections) is
shown in Fig.~\ref{fig:F}, according to which $F$ vanishes (and even becomes
negative, in which case we take $F=0$) for  $m_{A_1}
\gsim 8.8$~GeV. Correspondingly, the CLEO bounds on the NMSSM
parameters disappear for $m_{A_1} \gsim 8.8$~GeV. (For larger values of
$m_b$ as 5.3~GeV, $F$ would vanish only for $m_{A_1} \sim 9.4$~GeV $\sim
 m_\Upsilon$ as also indicated in Fig.~\ref{fig:F}.)

\begin{figure}[ht!]
\begin{center}
\includegraphics[scale=0.5,clip=,]{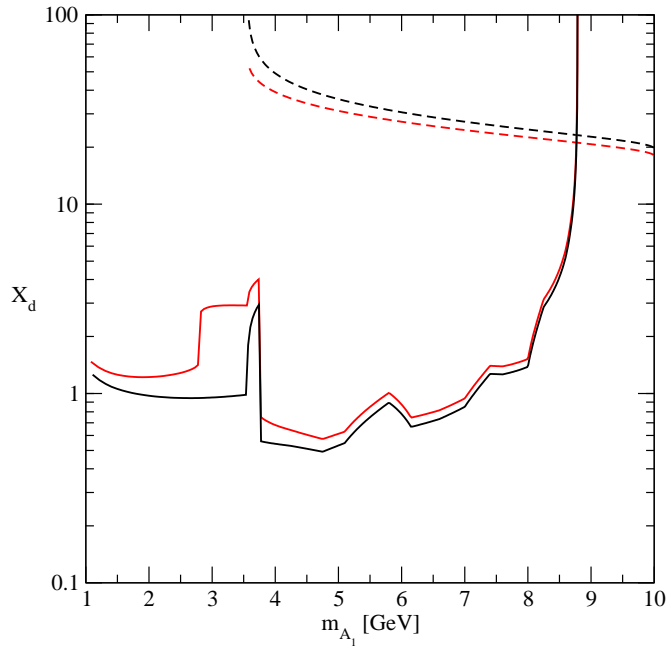}
\caption{Upper bounds on $X_d$ as a function of $m_{A_1}$ for two
extreme values of $\tan\beta= 1.5$ (red curve) and $\tan\beta= 50$
(black curve) using results from
CLEO~\cite{cleo}. We also indicate as dashed lines the region at large
$X_d$ and $m_{A_1} >  3.5$~GeV  where $\Gamma_{A_1}$  exceeds
10~MeV (same colour code for $\tan\beta= 1.5$ and $50$).}
\label{fig:cleo}
\end{center}
\end{figure}

Next, in order to translate the CLEO bounds into bounds on 
$X_d(m_{A_1})$ using eq. (\ref{eq:wilczek}), the branching ratios ${\cal
B} \left(A_1 \to \tau^+\tau^-/\mu^+\mu^-\right)$ have to be known, which
depend essentially on $\tan\beta$. For $m_{A_1}$ above 2\,$m_\tau$,
${\cal B} \left(A_1 \to \tau^+\tau^-\right)$ varies from $\sim 70\%$ for
$\tan\beta = 1.5$ to $\sim 95\%$ for $\tan\beta = 50$,
whereas ${\cal B} \left(A_1 \to \mu^+\mu^-\right)$ is
always below 10\% even for $m_{A_1}$ below 2\,$m_\tau$ (which implies
to reconsider the estimates of the CLEO reach in~\cite{McKeen:2008gd}).
Using the code NMSSMTools~\cite{nmssmtools} for the determination of the
${\cal B} \left(A_1 \to \tau^+\tau^-/\mu^+\mu^-\right)$ and an
interpolation of the CLEO bounds~\cite{cleo}, we show our resulting
upper limits on $X_d$ as a function of $m_{A_1}$ for two extreme values
of $\tan\beta= 1.5$ and 50 in Fig.~\ref{fig:cleo}.

Actually, in Ref.~\cite{cleo} the total decay width of $A_1$, $\Gamma_{A_1}$,
is assumed to be below $10$~MeV. Although we do not believe that the CLEO
bounds disappear completely in the case where $\Gamma_{A_1}$ (which
increases with $X_d$ and $m_{A_1}$) is larger than 10~MeV, we indicate
in Fig.~\ref{fig:cleo} also the region at large $X_d$ and $m_{A_1} \gsim
3.5$~GeV where $\Gamma_{A_1}$ exceeds 10~MeV (depending also
slightly on $\tan\beta$). In the updated version 2.1 of the
NMSSMTools package~\cite{nmssmtools} these bounds are included.

\section{Mixing of $A_1$ with the $\eta_b(nS)$ resonances}
\label{sec:mix}

In the presence of a pseudoscalar Higgs boson with a mass close to one
of the different $\eta_b(nS)$ resonances, a significant mixing  between
these states can occur~\cite{Drees:1989du}.

The mixing between a CP-odd Higgs and a single $\eta_b(nS)$ ($n=1,2$ or
3) resonance can be described by the introduction of off-diagonal
elements denoted by $\delta m_n^2$ in the mass
matrix~\cite{Drees:1989du,Fullana:2007uq} (here and below we neglect
possible induced $\eta_b(nS)-\eta_b(n'S)$ mixings for $n \neq n'$)
\beq\label{eq:massmatr}
{\cal M}_n^2=  
\left(
     \begin{array}{cc}
      m_{A_{10}}^2-im_{A_{10}}\Gamma_{A_{10}} & \delta m_n^2\\
      \delta m_n^2 &
      m_{\eta_{b0}(nS)}^2-im_{\eta_{b0}(nS)}\Gamma_{\eta_{b0}(nS)}      
     \end{array}
\right)
\end{equation} 
where the subindex '0' indicates unmixed states: $m_{A_{10}}$ and
$m_{\eta_{b0}(nS)}$ ($\Gamma_{A_{10}}$ and $\Gamma_{\eta_{b0}(nS)}$) 
denote the masses (widths) of the pseudoscalar Higgs boson and
$\eta_{b0}(nS)$ states, respectively, before mixing. 

In Ref.\cite{Fullana:2007uq} only the mixing of the Higgs with the
$\eta_{b0}(1S)$ resonance was taken into account. In this paper, we
extend the analysis by considering the possible mixing between the
Higgs and {\em any} of the three $\eta_{b0}(nS)\ (n=1,2$ or 3)
states. Thus three mixing angles have to be defined;
however, only the contribution  from the closest $\eta_b(nS)$ state to
the hypothetical $A_1$ mass will be assumed to be significant for the
mixing, i.e. only one among the three mixing angles will deviate
significantly from zero. The generally complex mixing angle $\alpha_n$
between the pseudoscalar Higgs $A_{10}$ and an $\eta_{b0}(nS)$ state is
given
by~\cite{Fullana:2007uq}
\begin{equation}\label{eq:alpha}
\sin 2\alpha_n\ =\ \delta m_n^2/\Delta_n^2
\end{equation}
where
\begin{equation} 
\Delta_n^2= [D_n^2+(\delta m_n^2)^2]^{1/2}
\end{equation}
with
\beq 
D_n=(m_{A_{10}}^2-m_{\eta_{b0}(nS)}^2-im_{A_{10}}\Gamma_{A_{10}}+
im_{\eta_{b0}(nS)}\Gamma_{\eta_{b0}(nS)})/2 \; .
\eeq
The off-diagonal element $\delta m_n^2$ can be computed
within the framework of a non-relativistic quark potential model as
\begin{equation}
\delta m_n^2\ =\  
\biggl(\frac{3m_{\eta_{b}(nS)}^3}{8 \pi v^2}\biggr)^{1/2}
|R_{\eta_b(nS)}(0)|
\times X_d\; .
\end{equation}

In a non-relativistic approximation to the bottomonium bound states, the
radial wave functions at the origin can be considered as identical for
vector and pseudocalar states,  i.e. $R_{\Upsilon(nS)}(0) \simeq
R_{\eta_b(nS)}(0)$, and can therefore be determined from the measured
$\Upsilon \to e^+e^-$ decay widths:
\begin{equation}\label{eq:leptonwidths}
\mid R_{\Upsilon(nS)}(0) \mid^2\ \simeq\  
\Gamma[\Upsilon(nS) \to e^+e^-]\times \frac{9m_{\Upsilon(nS)}^2}{4
\alpha^2} \biggl[1+\frac{16\alpha_s(m_{\Upsilon}^2)}{3\pi}\biggr]
\end{equation} 
Substituting recent values for the dielectron widths
from~\cite{Amsler:2008zz} we obtain $|R_{\eta_b(1S)}(0)|^2=6.60$
GeV$^3$, $|R_{\eta_b(2S)}(0)|^2=3.02$ GeV$^3$ and
$|R_{\eta_b(3S)}(0)|^2=2.18$ GeV$^3$, leading to\footnote{Similar values
can be obtained from a  Buchmuller-Tye potential~\cite{Eichten:1995ch}.}
\beq\label{dmest}
 \delta m_1^2\ =\ 0.14\ \mathrm{GeV}^2\times X_d,\quad
 \delta m_2^2\ =\ 0.11\ \mathrm{GeV}^2\times X_d,\quad
 \delta m_3^2\ =\ 0.10\ \mathrm{GeV}^2\times X_d\; .
\eeq 

The $A_1$ and $\eta_b(nS)$ physical (mixed) states can be written as
\begin{eqnarray}\label{eq:mix1}
A_1 &=& \cos\alpha_n\ A_{10}\ +\ \sin\alpha_n\ \eta_{b0}(nS)\; ,\nn
\\
\eta_b(nS) &=& \cos\alpha_n\ \eta_{b0}(nS)\ -\ \sin\alpha_n\ A_{10} 
\end{eqnarray}
assuming $\cos^2\alpha_n+\sin^2\alpha_n\simeq 1$, i.e.
neglecting the imaginary components of $\alpha_n$. (Here and below we
use the notation $A_1$ and $\eta_b(nS)$ for the mixed states in order to
indicate their dominant components for small mixing angles. Clearly, for
$\alpha_n \sim 90^o$, their dominant components would be reversed.)

The full widths $\Gamma_{A_{1}}$ and $\Gamma_{\eta_b(nS)}$ of the $A_1$
and $\eta_b(nS)$ physical states can be expressed in terms of the widths
of the unmixed states according to~\cite{Fullana:2007uq}
\begin{eqnarray}\label{eq:fullwid}
\Gamma_{A_1} &\simeq& 
\cos^2\alpha_n\ \Gamma_{A_{10}}\ +\  
\sin^2\alpha_n\ \Gamma_{\eta_{b0}(nS)}\; ,\nn
\\
\Gamma_{\eta_b} &\simeq&  
\cos^2\alpha_n\ \Gamma_{\eta_{b0}(nS)}\ +\  
\sin^2\alpha_n\ \Gamma_{A_{10}}\; .
\end{eqnarray}

Finally, let us recall that the mixing of the $A_{10}$ with $\eta_{b0}$
states should lead to mass shifts which can be
sizable~\cite{Drees:1989du, Fullana:2007uq}. These mass shifts might
have spectroscopic consequences concerning the hyperfine
$\eta_b(nS)-\Upsilon(nS)$ splitting~\cite{Drees:1989du,
SanchisLozano:2004gh, Fullana:2007uq} whose predictions within the SM
are reviewed in~\cite{Brambilla:2004wf}, and with respect to which the
BaBar result~\cite{Aubert:2008vj} on the $\eta_b(1S)-\Upsilon(1S)$
hyperfine splitting -- in the absence of a light CP-odd Higgs -- would
be somewhat large (see the next section).

\section{\bf Upper bounds on $X_d$ from the measured $\eta_b$
mass, and the\break mixing-induced $\eta_b$ mass shift}\label{sec:babar}

The observation of an $\eta_b$-like state with a mass of $\simeq
9.389$~GeV by BaBar~\cite{Aubert:2008vj}, allows to obtain upper limits
on the
reduced coupling $X_d$ as a function of the lightest CP-odd Higgs mass
parameter $m_{A_{10}}$, if $m_{A_{10}}$ is near 9.39 GeV. This follows
from the fact that the measured mass squared  has now to be considered
as (the real part of) the eigenvalue of the matrix ${\cal M}_1^2$
(\ref{eq:massmatr}), corresponding algebraic relations and an estimate
of hadronic parameters as $m_{\eta_{b0}(1S)}$.

Subsequently we denote the ``$\eta_b$'' mass as measured by
BaBar by $m_{obs}$, and the state $\eta_{b0}(1S)$ by $\eta_{b0}$.
The observed state has now to be considered as a superposition of
$A_{10}$ and $\eta_{b0}$. Then the following algebraic identity holds
(where $\delta m_1^2$ is the off-diagonal element of the matrix 
${\cal M}_1^2$ (\ref{eq:massmatr})):

\beq\label{algident}
\left(\delta m_1^2\right)^2 = \Delta_A \Delta_\eta \left[
1+\frac{\gamma^2}{\left(\Delta_A + \Delta_\eta\right)^2}\right]
\eeq
where
\beq\label{defdelta}
\Delta_A = m_{A_{10}}^2 - m_{obs}^2,\qquad \Delta_\eta = m_{\eta_{b0}}^2 -
m_{obs}^2
\eeq
and
\beq\label{defgam}
\gamma = m_{A_{10}}\Gamma_{A_{10}} - m_{\eta_{b0}}\Gamma_{\eta_{b0}}\; .
\eeq

Note that $\Delta_A$ and $\Delta_\eta$ must have the same sign, which
follows already from properties of eigenvalues of real $2\times 2$
matrices.

Now, if we use estimates for the parameters $m_{\eta_{b0}}$ and
$\gamma$, eq. (\ref{algident}) allows to obtain an upper bound on $X_d$
as a function of $m_{A_{10}}$. First, for $\gamma$ we can assume
$|\gamma| \lesssim m_{obs}\times 20$~MeV (from $\Gamma_{A_{10}}$,
$\Gamma_{\eta_{b0}} \lesssim 20$~MeV) with the result that the term
$\sim \gamma^2$ in (\ref{algident}) is relevant only for $m_{A_{10}}$
very close to $m_{obs}$.

For $(m_{A_{10}},\, m_{\eta_{b0}}) \sim m_{obs}$ (but $(m_{A_{10}},\,
m_{\eta_{b0}}) -  m_{obs}$ larger than a few~MeV such that the term $\sim
\gamma^2$ can be neglected), eq.~(\ref{algident}) can be simplified
further with the result
\beq\label{dmapp}
\left(\delta m_1^2\right)^2 \simeq 4 m_{obs}^2
\left(m_{A_{10}}-m_{obs}\right) \left(m_{\eta_{b0}} - m_{obs}\right)
\eeq
and hence, from (\ref{dmest}),
\beq\label{xdest}
X_d \simeq 125\ \sqrt{\left(m_{A_{10}}-m_{obs}\right) 
\left(m_{\eta_{b0}} - m_{obs}\right)}\ \mathrm{GeV}^{-1}\; .
\eeq

To proceed further, we have to estimate $m_{\eta_{b0}}$. Most of
previous estimates for $m_{\eta_{b0}}$ correspond actually to
$m_{\eta_{b0}} - m_{obs} > 0$~\cite{Godfrey:2001eb,Brambilla:2004wf},
but subsequently we allow for
\beq\label{asseta0}
m_{\eta_{b0}} - m_{obs} = -30\ \dots\ +40\ \mathrm{MeV}\; .
\eeq

Next we have to treat the cases $m_{A_{10}}-m_{obs} > 0$ and
$m_{A_{10}}-m_{obs} < 0$ separately. Starting with $m_{A_{10}}-m_{obs} <
0$, the maximally possible value for $X_d$ from (\ref{xdest}) is assumed
for the lowest estimate of $m_{\eta_{b0}}$, with the result
\beq\label{xdmaxm}
X_d^{max}(m_{A_{10}}) \sim 22 \sqrt{m_{obs}-m_{A_{10}}}\
\mathrm{GeV}^{-1/2}\; . \eeq
For $m_{A_{10}}-m_{obs} > 0$, on the other hand, the maximally possible
value for $X_d$ is assumed for the largest estimate of $m_{\eta_{b0}}$,
with the result
\beq\label{xdmaxp}
X_d^{max}(m_{A_{10}}) \sim 25 \sqrt{m_{A_{10}}-m_{obs}}\
\mathrm{GeV}^{-1/2}\; .
\eeq

These analytic expressions for $X_d^{max}(m_{A_{10}})$ are fairly good
approximations to the numerical upper bounds on $X_d(m_{A_{10}})$ which
can be derived from (\ref{algident}) without the approximation
(\ref{dmapp}), apart from the region where $|m_{A_{10}}-m_{obs}|$ is
less than about 0.5~MeV (where (\ref{xdmaxm}) and (\ref{xdmaxp}) would
imply $X_d^{max}(m_{A_{10}}) \to 0$). In fact, with $|\gamma| \sim
m_{obs}\times 20$~MeV, one obtains $X_d^{max}(m_{A_{10}}) \sim 0.6$ for
$|m_{A_{10}}-m_{obs}| \lsim 0.5$~MeV (see Fig.~\ref{fig:xd_ma1} below).

We emphasize, however, that most previous estimates for $m_{\eta_{b0}}$
correspond to $m_{\eta_{b0}} - m_{obs} >$~0 \cite{Godfrey:2001eb,
Brambilla:2004wf} in contrast to our more conservative assumption
(\ref{asseta0}). These estimates can still be correct within the present
framework, if an additional $A_{10}$ state with $m_{A_{10}}-m_{obs} > 0$
exists, which mixes strongly with the $\eta_{b0}$, reducing the lower
eigenvalue of the mass matrix (\ref{eq:massmatr}). The induced mass
shift $m_{\eta_{b0}} - m_{obs}$ can easily be derived from
eq. (\ref{xdest}):
\beq\label{eq:massshift}
m_{\eta_{b0}} - m_{obs} \simeq \frac{X_d^2\ \times 1\ \rm{GeV}^2} {1.56\cdot 10^4
\cdot \left(m_{A_{10}}-m_{obs}\right)}
\eeq
which, for $m_{\eta_{b0}} - m_{obs} >$~0, would unwittingly be interpreted
as an excess of the ``observed'' hyperfine splitting $m_{\Upsilon(1S)} -
m_{obs}$. For instance, an induced mass shift of $m_{\eta_{b0}} - m_{obs}
\sim 20$~MeV would be generated by a CP-odd Higgs with mass $m_{A_{10}}$
and a reduced coupling $X_d$ satisfying $X_d \simeq 17.7 \times
\sqrt{m_{A_{10}}-m_{obs}}$~GeV$^{-1/2}$ as, e.g., $X_d \simeq 12$ for 
$m_{A_{10}} \simeq 9.85$~GeV.

Note, however, that this mecanism would imply a heavier mass eigenvalue
of the mixing matrix (\ref{eq:massmatr}) is not too far above $m_{obs}$.
This favours a heavy mass eigenvalue below 10.5~GeV, which not only
satisfies LEP constraints but could even, as mentionned in the
introduction, explain an observed excess of events at LEP.

\section{\bf Comparison of constraints from CLEO, BaBar, B~physics and
the muon anomalous magnetic moment}\label{sec:comp}

In addition to the constraints obtained in section \ref{sec:cleo} from
CLEO and in section \ref{sec:babar} from BaBar, the
$(m_{A_1},X_d)$-plane is already constrained by processes from B
physics~\cite{Hiller:2004ii,Domingo:2007dx} and the muon anomalous
magnetic moment $(g-2)_\mu$~\cite{mumagmo}. (Upper limits on $X_d$ have
been derived by OPAL~\cite{opal} from Yukawa production of a light
neutral Higgs Boson at LEP, which seem more restrictive than the
constraints from CLEO for $m_{A_1} \gsim 9.2$~GeV. We believe, however,
that the $\eta_b(nS) - A_1$ mixing, which is relevant here, depends on
an additional $b$-$b$-$\eta_b$ form factor, where the initial $b$-quark
is far off-shell. Since this effect has not been considered
in~\cite{opal}, we will not consider the corresponding limits below.) 

In the following, we will compare the different constraints in the
$(X_d,m_{A_1})$- and $(X_d,M_{A})$-planes. (In this section, $m_{A_1}$
is the CP-odd Higgs mass parameter denoted as $m_{A_{10}}$ in the mass
matrix (\ref{eq:massmatr}). However, the difference between 
$m_{A_{10}}$ and $m_{A_1}$ would hardly be visible in the Figures
below.)

For this purpose we have performed a scan over the NMSSM parameter space
using the NMHDECAY program from the NMSSMTools
package~\cite{nmssmtools}. NMHDECAY allows to verify simultaneously
the phenomenological constraints from SUSY searches, Higgs searches,
B~physics and $(g-2)_\mu$. 
We have varied the NMSSM parameters (\ref{eq:inp}) $\lambda$, $\kappa$,
$A_\lambda$, $A_\kappa$, $\mueff$, $\tan\beta$ (the latter between 1
and 50), as well as the SUSY breaking gaugino, squark and slepton masses
and trilinear couplings, keeping only points where $m_{A_1} <
10.5$~GeV. Then we identified regions in the parameter space which are
ruled out by the various phenomenological constraints for {\it any} choice of
parameters. In particular, LEP constraints from  Higgs searches require
$\tan\beta \gsim 1.5$ in the NMSSM, while constraints from $(g-2)_\mu$
lead to $\tan\beta \gsim 2$ for $m_{A_1} < 10.5$~GeV.

\begin{figure}[ht!]
\begin{center}
\includegraphics[scale=0.5,clip=,]{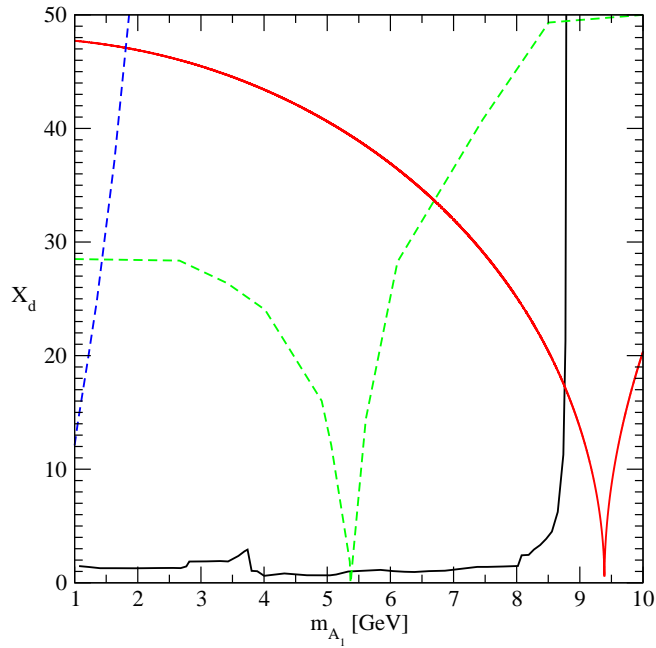}
\caption{Upper bounds on $X_d$ versus the $A_1$ mass for all parameters
scanned over (see text). Indicated are constraints from $B_s \to
\mu^+\mu^-$ and $\Delta M_q$, $q=d,s$ as a green dashed line,
constraints from $(g-2)_\mu$ as a blue dashed line, the latest bounds from
CLEO on ${\cal B}\left(\Upsilon \to \gamma \tau \tau\right)$ as a black
line and constraints due to the measured $\eta_b(1S)$ mass by
Babar as a red line.}
\vspace*{-5mm}
\label{fig:xd_ma1}
\end{center}
\end{figure}

The various curves in the $(X_d,m_{A_1})$-plane in Fig.~\ref{fig:xd_ma1}
indicate lower bounds on $X_d$ from various phenomenological constraints.
We found that even for very large $X_d$ there always exist parameter choices
such that no region is {\it always} excluded by either the constraints from 
${\cal B}\left(B\to X_s\gamma\right)$ or ${\cal B}\left(\bar{B}^+
\to \tau^+\nu_{\tau}\right)$. However, constraints from ${\cal
B}\left(B_s \to \mu^+\mu^-\right)$ and $\Delta M_q$, $q=d,s$ (shown as a
green dashed line) always
exclude a funnel for $m_{A_1}\sim M_{B_q}\sim 5.3$~GeV, the width of
which depends on the loop-induced $b-s-A_1$ coupling. This coupling
being proportional to $X_d$, the excluded region broadens steadily with
$X_d$ as can be observed in Fig.~\ref{fig:xd_ma1}, leading to the
exclusion of all pseudoscalars with masses below $\sim 6$~GeV for
$X_d\gsim 30$. However, the CLEO constraints indicated as a black line
are much more restrictive, apart from a narrow window around
$m_{A_1}\sim 5.3$~GeV.

Constraints from $(g-2)_\mu$ originate
from the contribution of a light pseudoscalar, which is enhanced  by
$X_d^2$. For $m_{A_1}$ below $\sim 3$~GeV, the pseudoscalar contribution
has the opposite (negative) sign
with respect to the deviation of the  Standard Model prediction from the
measured value of $(g-2)_{\mu}$~\cite{mumagmo}. This results in the
exclusion of very light $A_1$ below $\sim 2$~GeV for large values of
$X_d$, as indicated in Fig.~\ref{fig:xd_ma1} -- a region which is now
also covered by CLEO constraints.

Finally, constraints due to the measured $\eta_b(1S)$ mass by Babar as
discussed in section~\ref{sec:babar} exclude a funnel around
$m_{A_1}\sim 9.4$~GeV, which is outside the region
covered by CLEO.

\begin{figure}[ht!]
\begin{center}
\vspace*{1cm}
\includegraphics[scale=0.5,clip=,]{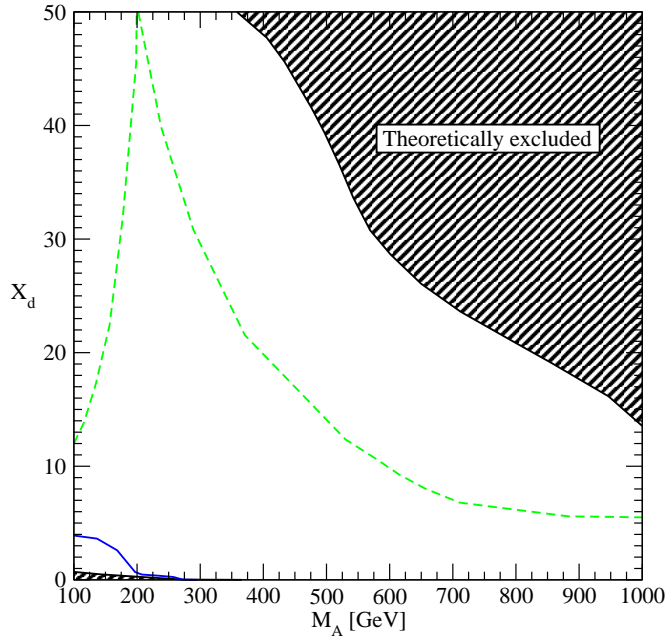}
\caption{Bounds on $X_d$ versus $M_A$ for $A_1$ masses
below 10.5~GeV. Indicated are upper bounds on $X_d$ from $B_s \to
\mu^+\mu^-$ and $\Delta M_q$, $q=d,s$ as a green dashed line,
and lower bounds on $X_d$ (for $m_A \lesssim 200$~GeV) from ${\cal
B}\left(B\to X_s\gamma\right)$ as a blue line.}
\label{fig:ma_xd}
\end{center}
\end{figure}

In the $(X_d,M_A)$-plane shown in Fig.~\ref{fig:ma_xd} one sees that, as
discussed qualitatively in section \ref{sec:nmssm}, large values of
$X_d$ can occur only for not too large values of $M_A$; here this
statement can be verified quantitatively. No region is generically
excluded by the constraints from CLEO, Babar or $(g-2)_\mu$, since
$m_{A_1}$ varies from 1 to 10.5~GeV for each point in this plane. The
upper bounds on $X_d$ from $B_s \to \mu^+\mu^-$ and $\Delta M_q$,
$q=d,s$ are indicated as in
Fig.~\ref{fig:xd_ma1}, and lower bounds on $X_d$ from
${\cal B}\left(B\to X_s\gamma\right)$ now appear as well at small $M_A$.
(There, the contribution to ${\cal B}\left(B\to X_s\gamma\right)$ from a
charged Higgs with a mass $\sim M_A$ has to be compensated by a
contribution $\sim X_d$ involving charginos or neutralinos, which
requires a sufficiently large value for $X_d$.)

The most important results of this section are contained in 
Fig.~\ref{fig:xd_ma1}, which shows that the combined present constraints
rule out most of the region where $m_{A_1} \lsim 8.5$~GeV -- except if
$X_d$ is sufficiently small -- whereas $m_{A_1} \sim 8.5-10.5$~GeV
remains an interesting region in parameter space allowing for large
$X_d$. The corresponding necessary values of $M_A$ can be deduced from
Fig.~\ref{fig:ma_xd}.

\section{Possible lepton universality breaking}\label{sec:luv}

As pointed out in~\cite{SanchisLozano:2002pm}, one
manifestion of the existence of a light CP-odd Higgs boson of mass
around $\sim 10$~GeV could be a breakdown of LU (lepton universality)
in $\Upsilon$ decays, if the (not necessarily soft) radiated photon
escapes undetected in the experiment, or simply is not specifically
searched for in the analysis of events. (The leptonic width is, in fact,
an inclusive quantity with a sum over an infinite number of photons.)
Higgs-mediated $\Upsilon$ decays would lead to an excess of its tauonic
branching ratio (BR), which can be assessed through the ratio
\begin{equation}
{\cal R}_{\tau/\ell}= \frac{{\cal B}_{\tau\tau}-{\cal B}_{\ell\ell}}
{{\cal B}_{\ell\ell}}= \frac{{\cal B}_{\tau\tau}}{{\cal B}_{\ell\ell}}-1
\label{eq:R}
\end{equation}
where ${\cal B}_{\tau\tau}$ denotes the tauonic, and ${\cal
B}_{\ell\ell}$ the electronic ($\ell=e$) or muonic ($\ell=\mu$)
branching ratios of the $\Upsilon$ resonance, respectively. A
statistically significant non-zero value of ${\cal R}_{\tau/\ell}$ would
be a strong argument in favour of a pseudoscalar Higgs boson mediating
the process.

In Table~\ref{FACTORES} we summarize the current situation of LU
obtained from~\cite{Amsler:2008zz}. As already mentionned in the
introduction, a $\sim 1\,\sigma$ effect seems visible in most cases,
leading to an overall (positive) $\sim 2\,\sigma$ effect.

\begin{table*}[hbt!]
\begin{center}
\begin{tabular}{|c|c|c|c|c|c|}
\hline
 & ${\cal B}\left(e^+e^-\right)$ & ${\cal B}\left(\mu^+\mu^-\right)$ &
 ${\cal B}\left(\tau^+\tau^-\right)$ & $R_{\tau/e}(nS)$ &
$R_{\tau/\mu}(nS)$\\ \hline
$\Upsilon(1S)$ & $2.38 \pm 0.11$ & $2.48 \pm 0.05$ & $2.60 \pm 0.10$ &
$0.09 \pm 0.06$ & $0.05 \pm 0.04$ \\ \hline
$\Upsilon(2S)$ & $1.91 \pm 0.16$ &  $1.93 \pm 0.17$ & $2.00 \pm 0.21$ &
$0.05 \pm 0.14$ & $0.04 \pm 0.06$ \\ \hline
$\Upsilon(3S)$ & $2.18 \pm 0.21$ & $2.18 \pm 0.21$ & $2.29 \pm 0.30$ &
$0.05 \pm 0.16$ & $0.05 \pm 0.16$ \\ \hline
\end{tabular}
\end{center}
\caption{Measured leptonic branching ratios ${\cal B}\left(\Upsilon(nS)
\to \ell \ell\right)$ (in \%) and error bars (summed in quadrature) of
$\Upsilon(1S)$, $\Upsilon(2S)$, and $\Upsilon(3S)$ resonances.} 
\label{FACTORES}
\end{table*}

Subsequently we intend to estimate the possible amount of LU breaking
in the NMSSM with $m_{A_1}$ in the 9 -- 10.5~GeV range and large $X_d$
in order to verify whether it can be assessed
experimentally~\cite{Fullana:2007uq}, i.e. whether it can be of the
order of the few percent.

In principle, also pure SM channels may yield an apparent breaking of LU
in $\Upsilon$ decays. In fact, all decays through an intermediate
pseudoscalar (e.g. $\eta_b$) state should break LU due to the leptonic
mass dependence of the amplitude as a consequence of helicity
conservation. However, such processes (mediated by a two-photon loop or
a $Z^0$-boson in the SM) provide contributions to the branching
ratios well below 1\% and can be dropped. On the other hand, phase
space should {\em suppress} the tauonic BR (by about 0.5\%) with
respect to the electronic and muonic modes.

\begin{figure}[ht!]
\begin{center}
\includegraphics[width=13pc]{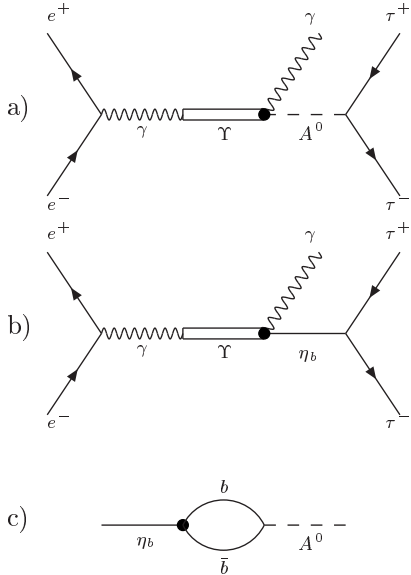}
\end{center}
\caption{Process $e^+e^-\to \Upsilon \to \gamma\ \tau^+\tau^-$  with a)
pseudoscalar Higgs, b) $\eta_b$ (after mixing) as intermediate states;
c) mixing diagram. Diagrams similar to a) and b) could be drawn for the
two-gluon decay mode yielding hadrons in the final state instead of
taus.}
\label{fig:feynman}
\end{figure}

In the absence of mixing between the Higgs boson $A_1$ and the $\eta_b$
resonances, the relevant contribution to ${\cal R}_{\tau/\ell}$ would
originate exclusively from the first diagram~a) in
Fig.~\ref{fig:feynman}. Assuming a BR($A_1 \to \tau^+\tau^-$) of 100\%
(see, however, below), $R_{\tau/\ell}$ would be given by the Wilczek
formula~(\ref{eq:wilczek}):
\begin{equation}\label{eq:wilc2}
{\cal R}_{\tau/\ell}=
\frac{{\cal B} \left(\Upsilon(nS) \to \gamma A_1\right)}
{{\cal B} \left(\Upsilon(nS) \to \mu^+\mu^-\right)}\equiv
R_0= \frac{G_F m_b^2 X_d^2}
{\sqrt{2}\pi\alpha}\biggl(1-\frac{m_{A_1}^2}{m_{\Upsilon}^2}\biggr)
\times F\; .
\label{eq:wtree}
\end{equation}

We recall that here we are interested in $A_1$ masses above 9~GeV, in
which case the very conservative (small) estimate in
section~\ref{sec:cleo} of the correction factor $F$ in (\ref{eq:wilc2})
ceases to make sense. The aim of section~\ref{sec:cleo} was to derive
conservative upper bounds on $X_d$; here, however, we aim at
realistic estimates of ${\cal R}_{\tau/\ell}$. To this end, the
tree-level expression (\ref{eq:wilc2}) corrected by a constant factor
$F=1/2$ should provide an acceptable
approximation~\cite{haber,Hiller:2004ii}. Then, the CLEO bounds exclude large
values of $X_d$ for $m_{A_1}$ up to roughly $9.2$~GeV.

In the presence of mixing between the Higgs boson $A_1$ and one of the
the $\eta_b$ resonances as in section~\ref{sec:mix}, both eigenstates
defined in eqs.~(\ref{eq:mix1}) would contribute to ${\cal
R}_{\tau/\ell}$. In order to give the $\Upsilon$ branching ratios into
these mixed states, it is convenient to define the ratio
\begin{equation}\label{eq:defs0}
S_0(n,n')\equiv
\frac{{\cal B} \left(\Upsilon(nS) \to \gamma \eta_{b0}(n'S)\right)}
{{\cal B} \left(\Upsilon(nS) \to \mu^+\mu^-\right)}\; ,
\end{equation}
where the branching ratio for a M1 transition between $\Upsilon(nS)$ and
$\eta_{b0}(n')$ states ($n'\leq n$) is given by~\cite{Godfrey:2001eb}
\begin{equation}\label{eq:m1trans}
{\cal B}\left(\Upsilon(nS) \to \gamma \eta_{b0}(n'S)\right)
=\frac{16\alpha}{3}\biggl(\frac{Q_b}{2m_b}\biggr)^2\ 
\frac{I_{n'n}^2 \cdot k^3}{\Gamma_{\Upsilon}}\; .
\end{equation}

$k$ is the photon energy (depending on the mass difference
$m_{\Upsilon(nS)}-m_{\eta_b(n'S)}$); $I_{n'n}$ denotes the final and
initial wave functions overlap, $I_{n'n}=\langle
f_{n'}|j_0(kr/2)|i_n\rangle$, where $j_0$ is a spherical Bessel
function. $I_{n'n}$ is numerically close to unity for favoured
transitions ($n=n'$) but much smaller for hindered ($n \neq n'$)
transitions. As stressed in~\cite{Godfrey:2001eb}, however, the
considerably larger photon energy $k$ in the latter case could
compensate  this reduction, leading to competitive transition
probabilities. Below we set $I_{12}=0.057$~\cite{Godfrey:2001eb} and
$I_{13}=0.017$. (The latter value is required  in order to reproduce the
experimental value ${\cal B}\left(\Upsilon(3S) \to \gamma\ {\rm
hadrons}\right)=4.8 \times 10 ^{-4}$  found by
BaBar~\cite{Aubert:2008vj}.) 

In terms of $R_0$ and $S_0$ defined above, the $\Upsilon$ branching
ratios into the mixed states $A_1$ and $\eta_b(n'S)$ are given by
(neglecting interference terms, and normalized w.r.t. the branching
ratios into $l^+l^-$ = $\mu^+ \mu^-$ or $e^+ e^-$)
\bea\label{eq:upstomixed}
\frac{{\cal B} \left(\Upsilon(nS) \to \gamma A_{1}\right)}
{{\cal B} \left(\Upsilon(nS) \to l^+l^-\right)}
&=& \cos^2 \alpha_k\ R_0 + \sin^2 \alpha_k\ S_0(n,k)\nn \\
\frac{{\cal B} \left(\Upsilon(nS) \to \gamma \eta_{b}(n'S)\right)}
{{\cal B} \left(\Upsilon(nS) \to l^+l^-\right)}
&=& \sin^2 \alpha_{n'}\ R_0 + \cos^2 \alpha_{n'}\ S_0(n,n')
\eea
where, as stated above, we assume that at most one possible mixing angle
$\alpha_k$ is nonvanishing. (For a given value for $m_{A_{10}}$, the
index $k$
is given by the state $\eta_{b0}(kS)$ whose mass is closest to 
$m_{A_{10}}$. The $\Upsilon$ decays into the remaining unmixed
$\eta_b(n'S)$ states with $n'\neq k$ are still described by eq.
(\ref{eq:m1trans}).)

Next we assume that the mixed states $A_1$ and $\eta_b(nS)$ decay
into $\tau^+ \tau^-$ only via their $A_{10}$ component.
Then we obtain,
using eq.~(\ref{eq:fullwid}) for the full widths of the mixed states,
\bea\label{eq:mixedbrs}
{\cal B} \left(A_1 \to \tau^+ \tau^-\right) &=& 
{\cal B} \left(A_{10} \to \tau^+ \tau^-\right) \times
\frac{\cos^2\alpha_k\ \Gamma_{A_{10}}}
{\cos^2\alpha_k\ \Gamma_{A_{10}}+\sin^2\alpha_k\
\Gamma_{\eta_{b0}(kS)}}\; ,\nn \\
{\cal B} \left(\eta_{b}(nS) \to \tau^+ \tau^-\right) &=& 
{\cal B} \left(A_{10} \to \tau^+ \tau^-\right) \times
\frac{\sin^2\alpha_n\ \Gamma_{A_{10}}}
{\cos^2\alpha_n\ \Gamma_{\eta_{b0}(nS)}+\sin^2\alpha_n\
\Gamma_{A_{10}}}\; ,
\eea
i.e. ${\cal B} \left(\eta_{b}(nS) \to \tau^+ \tau^-\right)$ vanishes for
$n \neq k$.
Finally we obtain for ${\cal R}_{\tau/\ell}$ (for a given
$\Upsilon(nS)$, and assuming ${\cal B} \left(A_{10} \to \tau^+
\tau^-\right) = 90\%$)
\bea\label{eq:rtl}
{\cal R}_{\tau/\ell} &=& {\cal R}_{\tau/\ell}^{A_1} +
{\cal R}_{\tau/\ell}^{\eta_b} \nn \\
&\equiv& \frac{{\cal B} \left(\Upsilon(nS) \to \gamma
A_{1}\right)} {{\cal B} \left(\Upsilon(nS) \to l^+l^-\right)}
\times {\cal B} \left(A_1 \to \tau^+ \tau^-\right)
+ \frac{{\cal B} \left(\Upsilon(nS) \to \gamma \eta_{b}(kS)\right)}
{{\cal B} \left(\Upsilon(nS) \to l^+l^-\right)}
\times {\cal B} \left(\eta_{b}(kS) \to \tau^+ \tau^-\right)\nn \\
&=& 0.9\ \left(\cos^2 \alpha_k\ R_0 + \sin^2 \alpha_k\
S_0(n,k)\right) \times \frac{\cos^2\alpha_k\ \Gamma_{A_{10}}}
{\cos^2\alpha_k\ \Gamma_{A_{10}}+\sin^2\alpha_k\
\Gamma_{\eta_{b0}(kS)}}\nn \\
&+& 0.9\ \left(\sin^2 \alpha_k\ R_0 + \cos^2 \alpha_k\
S_0(n,k)\right) \times \frac{\sin^2\alpha_k\ \Gamma_{A_{10}}}
{\cos^2\alpha_k\ \Gamma_{\eta_{b0}(kS)}+\sin^2\alpha_k\
\Gamma_{A_{10}}}
\eea
where either $\alpha_k =0$ (if none of the states $\eta_b(kS)$ mixes
with $A_{10}$), or $\alpha_k$ is given by the (supposedly only) mixing
angle, whose choice and value depend on $m_{A_{10}}$.

\begin{figure}[ht!]
\begin{center}
\includegraphics[width=16pc]{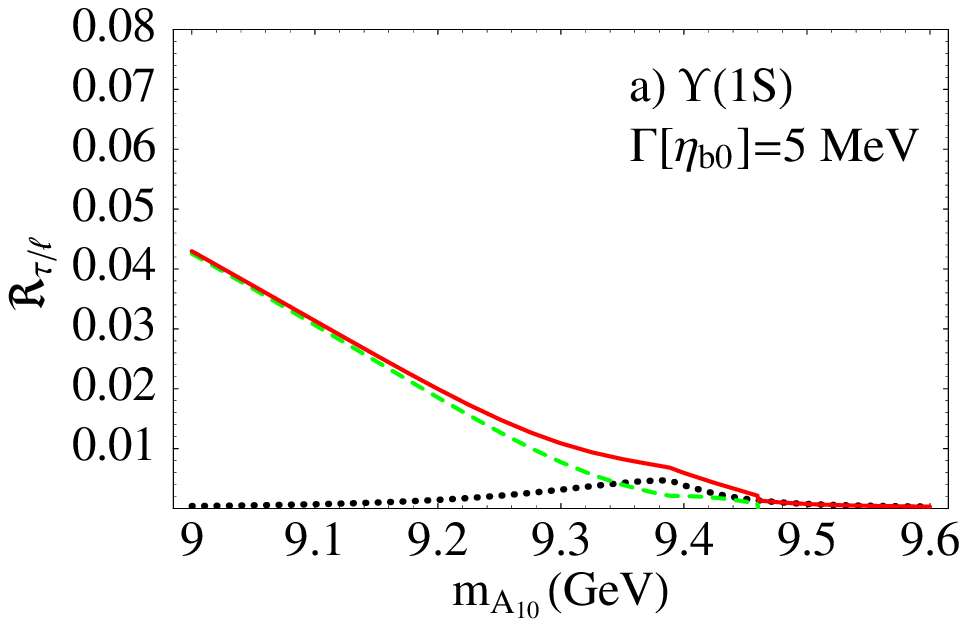}
\includegraphics[width=16pc]{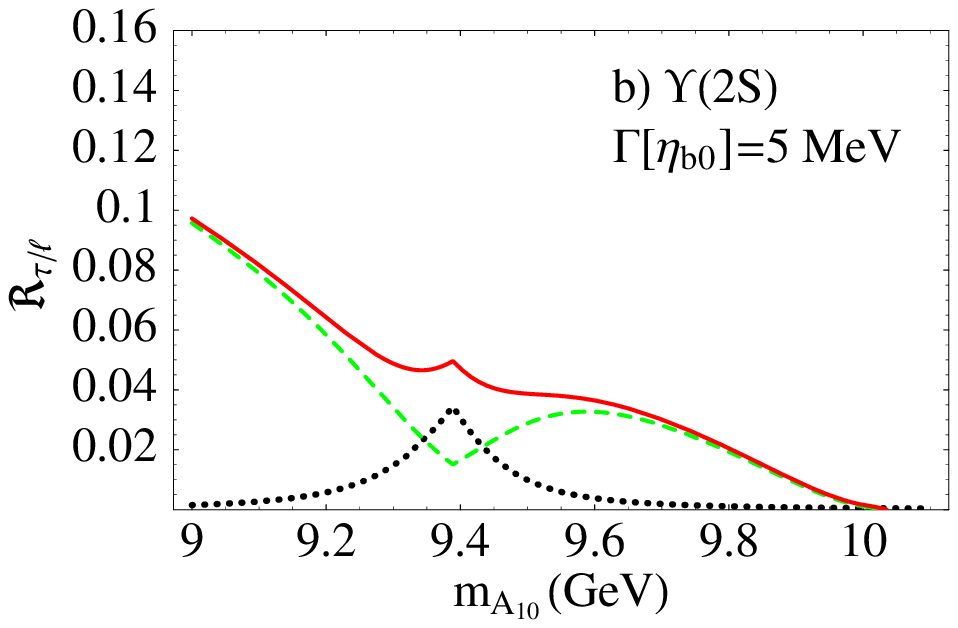}
\includegraphics[width=16pc]{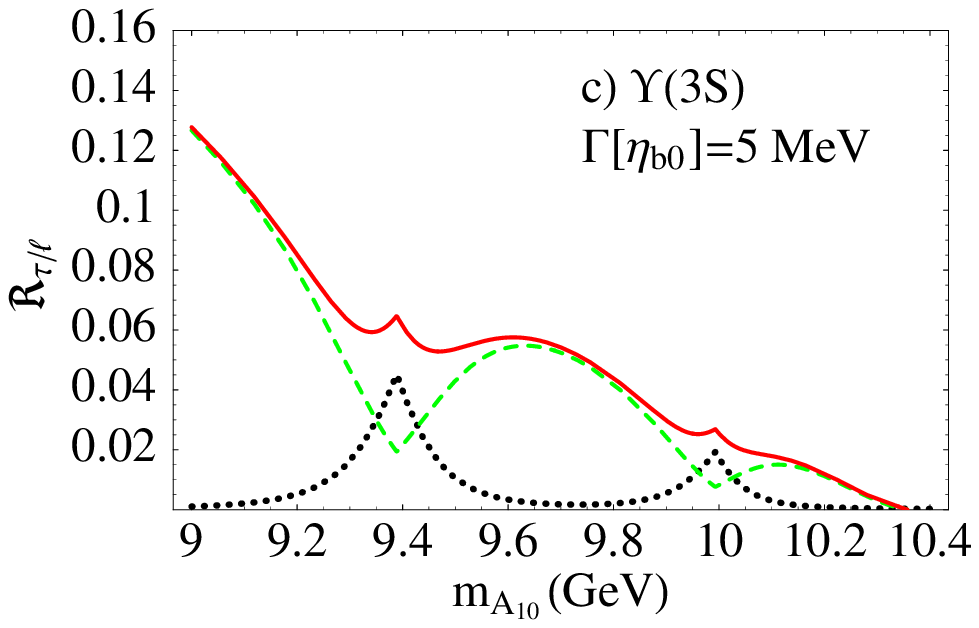}
\end{center}
\caption{${\cal R}_{\tau/\ell}$ versus the pseudoscalar Higgs mass for
$a)$ $\Upsilon(1S)$, $b)$ $\Upsilon(2S)$, and $c)$ $\Upsilon(3S)$ decays 
using $X_d=12$, $m_{\eta_{b0}(1S)}=9.389$ GeV~\cite{Aubert:2008vj},
assuming $m_{\eta_{b0}(2S,3S)}=9.997,10.32$ GeV respectively,  and
$\Gamma_{\eta_{b0}(1S,2S,3S)}=5$ MeV. The contributions from ${\cal
R}_{\tau/\ell}^{A_1}$ are indicated as dashed green lines, the
contributions from ${\cal R}_{\tau/\ell}^{\eta_b}$ as dotted black
lines, and their sum ${\cal R}_{\tau/\ell}$ as solid red lines. Larger
(smaller) values of $X_d$ obviously yield higher (lower) values
for $R_{\tau/\ell}$.}
\label{fig:r_all_s}       
\end{figure}

In the set of plots of Fig.~\ref{fig:r_all_s}, ${\cal R}_{\tau/\ell}$ is
shown for $\Upsilon(1S)$, $\Upsilon(2S)$ and $\Upsilon(3S)$,
respectively, as a function of $m_{A_{10}}$, using the formulas of
section~4 for the determination of the
relevant mixing angle. We assume tentatively\footnote{The following
expression (asymptotically valid for very heavy quark masses) can be
used to estimate the $\eta_b(nS)$ full width from the experimentally
known $\eta_c(nS)$ full width: $\Gamma_{\eta_b}(nS)/\Gamma_{\eta_c}(nS)
\simeq  (m_b/m_c)[\alpha_s(2m_b)/\alpha_s(2m_c)]^5$~\cite{oliver}. The
fifth power  of the $\alpha_s$ ratio yields however a large uncertainty
to the prediction. On the other hand, theoretical predicitions based on
the expected  ratio of the two-photon and two-gluon widths range from 4
to 20 MeV~\cite{Kim:2004rz}} $\Gamma_{\eta_{b0(1S)}.}
=\Gamma_{\eta_{b0(2S)}} =\Gamma_{\eta_{b0(3S)}}=5$~MeV and $X_d = 12$.
(As argued in section~5, values for $m_{A_1} \sim 9.4 \pm 0.2$~GeV are
actually ruled out for this value of $X_d$. However, for lower values of
$X_d$ -- leading to correspondingly lower values for ${\cal
R}_{\tau/\ell}$ -- the forbidden window for $m_{A_1}$ becomes smaller.
Moreover, phenomenologically interesting values for $m_{A_1}\gsim
9.6$~GeV as discussed at the end of section~5, are seen to generate
interesting values for  ${\cal R}_{\tau/\ell}$.)

The contributions from ${\cal R}_{\tau/\ell}^{\eta_b}$ (dotted black
line) yield the expected bumps around the respective $\eta_b$ mass
values, where the mixing angle becomes large. Conversely, the
contributions from ${\cal R}_{\tau/\ell}^{A_1}$ (dashed green line) show
dips at both $m_{A_1}=m_{\eta_b(1S)}=9.389$~GeV and
$m_{A_1}=m_{\eta_b(2S)} \simeq 10$~GeV, since they become reduced by the
mixing. (The expected peak or dip at $m_{A_1}=m_{\eta_b(2S)} \simeq
10.3$~GeV in Fig.~\ref{fig:r_all_s}c) is in fact invisibly small.)

\begin{figure}[ht!]
\begin{center}
\includegraphics[width=16pc]{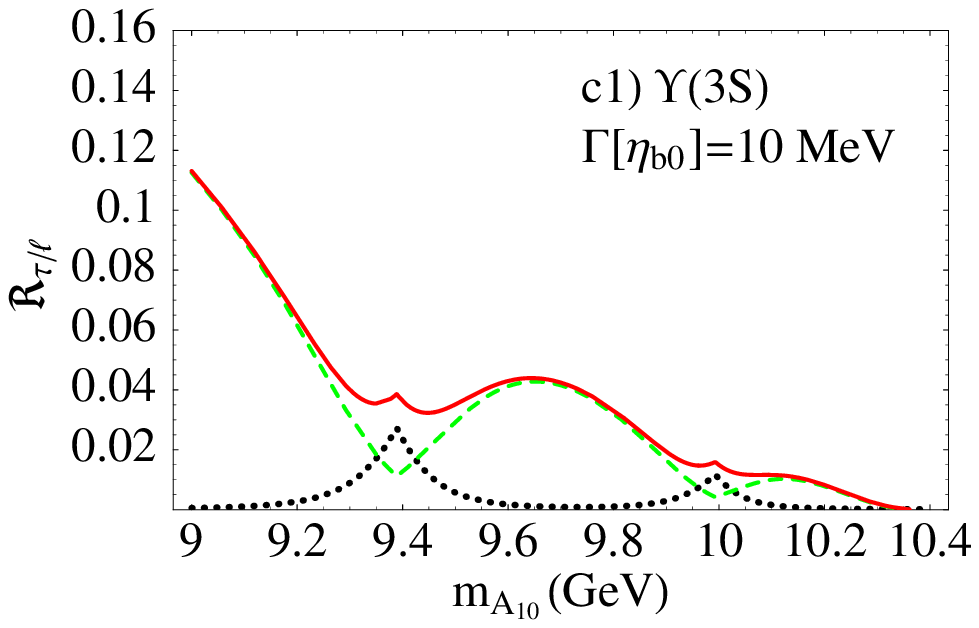}
\includegraphics[width=16pc]{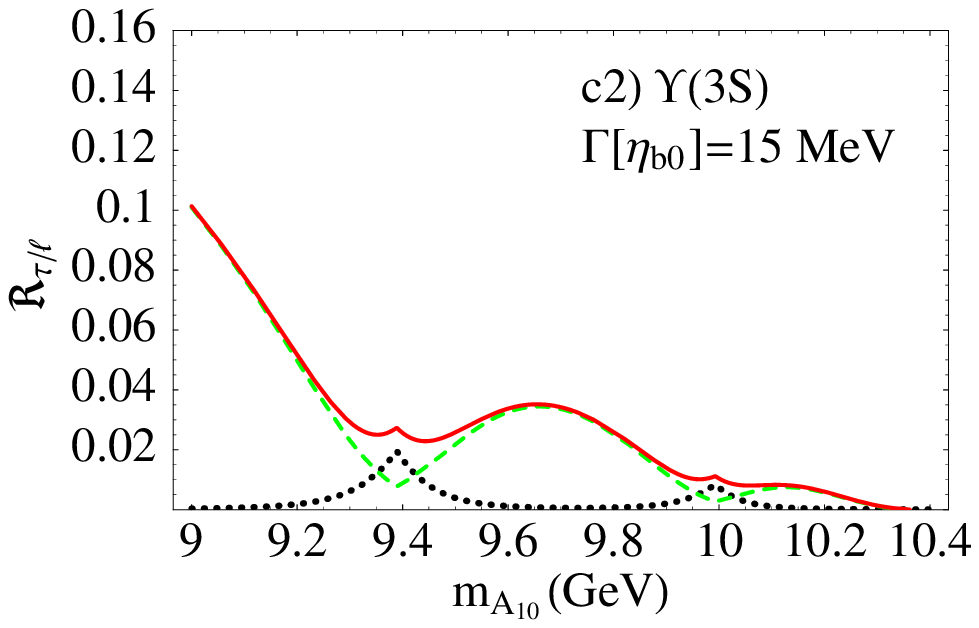}
\end{center}
\caption{$R_{\tau/\ell}$ versus the pseudoscalar Higgs  mass for
$\Upsilon(3S)$ decays using $X_d=12$ and $c1)$ $\Gamma_{\eta_{b0}}=10$
MeV, $c2)$  $\Gamma_{\eta_{b0}}=15$ MeV, respectively.}
\label{fig:gamma=10,15}       
\end{figure}

The higher values of ${\cal R}_{\tau/\ell}$ and the higher reach in
$m_{A_{10}}$ obtained for the $\Upsilon(2S)$ and $\Upsilon(3S)$ (due to
the dominant Wilczek mechanism of Fig.~\ref{fig:feynman}a)) allow us to
conclude that radiative decays of the latter resonances look more
promising than the $\Upsilon(1S)$ decays, allowing for the experimental
observation of LU breaking (at the few percent level) at a B factory.
This result is important for future tests of LU~\cite{Bona:2007qt}. 

In order to study the effect of our assumption on the width 
$\Gamma_{\eta_{b0}}$, we present in the set of
Fig.~\ref{fig:gamma=10,15} ${\cal R}_{\tau/\ell}$ for the $\Upsilon(3S)$
resonance setting $\Gamma_{\eta_{b0}}=$10 and 15 MeV, respectively. One
can observe a slight overall decrease of ${\cal R}_{\tau/\ell}$  for
larger $\Gamma_{\eta_{b0}}$, as expected from eqs.~(\ref{eq:mixedbrs})
and (\ref{eq:rtl}).

\begin{figure}[ht!]
\begin{center}
\includegraphics[width=25pc]{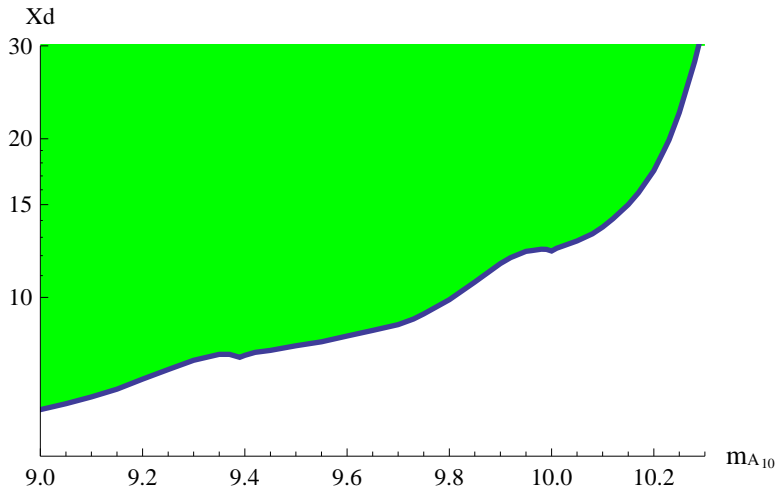}
\caption{Expected 2\ $\sigma$ signal (green area) in the mass range
$9-10.35$ GeV of the CP-odd Higgs $A_1$, 
assuming a total error of 2\% for $R_{\tau/\ell}$.}
\label{fig:foreseen}
\end{center}
\end{figure}

Concerning future measurements, we assume tentatively that a combined
statistical and systematic error (summed in quadrature) of 2\% for
$R_{\tau/\ell}$ is achievable at a (Super) B factory.
Fig.~\ref{fig:foreseen} shows the foreseen $2\ \sigma$ (95\% CL) limits 
(green region) for testing LU using $\Upsilon(3S)$ decays for a
$A_1$ mass ranging in the interval 9--10.3~GeV.

An observation of lepton universality breaking in $\Upsilon$ decays
should lead to a careful search for the quasi-monochromatic photons
shown in Figs.~\ref{fig:feynman} in a sample of events firstly selected
and enriched using 1-prong tauonic decays and requiring missing energy
(neutrinos). Let us recall that there could be two nearby peaks 
corresponding to two physical eigenstates. However, large $A_1$ or
$\eta_b$ widths might invalidate this search method: if the $m_{A_1}$
and $m_{\eta_b}$ masses were not too different (i.e. less than 50 MeV),
the two peaks might not be resolved experimentally but yield a broader
peak than expected. This conventional search has been unsuccessful so
far in $\Upsilon(1S)$ decays, but can (should) be extended to the yet
unexplored radiative decays of the $\Upsilon(2S,3S)$ resonance into
$\tau$'s, according to the proposal
in~\cite{Sanchis-Lozano:2006gx,Fullana:2007uq}.

Finally, a light CP-odd Higgs which mixes with one of the $\eta_b$
states would also decay hadronically and eventually be visible by
requiring four or more charged tracks together with a photon in
radiative $\Upsilon(2S,3S)$ decays, a criterium used by
Babar~\cite{Aubert:2008vj} for their discovery of the $\eta_b(1S)$.
Whereas the Babar data can possibly be used to put bounds on such an
additional state for certain ranges of its mass, we believe that a
serious search for such an additional state in the inclusive photon
spectrum from radiative $\Upsilon$ decays would require a detailed
treatment of the various background contributions, which is beyond the
scope of the present paper. Corresponding investigations are clearly
another interesting task in the future.

\section{Conclusions and outlook}\label{sec:concl}

In this paper we have summarized constraints from and perspectives of
various processes related to a CP-odd Higgs boson with a mass $m_{A_1}$
below the $B \bar{B}$ threshold of 10.5~GeV. Apart from $m_{A_1}$, these
phenomena depend essentially on its reduced coupling $X_d$ to
$b$-quarks. Within the parameter space of the NMSSM, relatively large
values of $X_d$ are possible (for sufficiently large values of
$\tan\beta$).

We have compared present constraints on the $m_{A_1} - X_d$ plane from
$B$ meson physics, the anomalous magnetic moment of the muon, LEP, and
recent results from CLEO and Babar. In spite of the conservative approach
towards the bound state corrections to the Wilczek formula the most
stringent constraints originate -- not astonishingly -- from the
dedicated (negative) searches by CLEO for $m_{A_1} \lsim 8.8$~GeV
allowing, however, for substantial values of $X_d$ 
provided that 8.8~GeV~$\lsim
m_{A_1} \lsim 10.5$~GeV.

Given the possible explanation of the 2.3 $\sigma$ excess in searches
for a CP-even Higgs boson at LEP~\cite{Dermisek:2005gg}, this allowed
mass range is of particular interest. We emphasize again that the
interval 9.4~GeV $\lsim m_{A_1} \lsim 10.5$~GeV can also have an effect on the
$\eta_b(1S)$ mass as measured by Babar via mixing, and explain the
possibly excessive $\Upsilon(1S)-\eta_b(1S)$ hyperfine splitting.

Such a scenario can and should be tested at presently
running $B$ factories, and/or a future Super $B$ factory. Obvious search
strategies consist -- as already performed -- in radiative
$\Upsilon(nS)$ decays into both tauonic and hadronic final states,
keeping an eye on possible close (but separate) peaks in the photon
spectrum. In addition, violation of lepton universality in inclusive
radiative $\Upsilon$ decays can be a signal for an additional CP-odd
Higgs. We have clarified that corresponding visible signals are well
within the reach of future precision experiments.

These searches are complementary to Higgs boson searches at colliders
(like the LHC) where it is quite doubtful at present whether a light
CP-odd Higgs decaying dominantly into $\tau^+\tau^-$ could be seen. On
the contrary, such a CP-odd Higgs can render searches for the lightest
CP-even Higgs boson $h$ very difficult, if it decays dominantly into $h
\to A_1 A_1 \to \tau^+\tau^-\tau^+\tau^-$. Hence, searches at $B$
factories are possibly our only windows into the light
Higgs sector, if such a scenario is realized.

\section*{Acknowledgments}

F.D. and U.E. gratefully acknowledge discussions with E. Kou.
This work was supported in part by the research grants FPA2005-01678
and FPA2008-02878, and in part by the U.S. Department of Energy,
Division of High Energy Physics, under Contract DE-AC02-06CH11357.
M.-A.S.-L. thanks MICINN for partial support.

\end{document}